\documentclass{ws-procs9x6}

\begin{document}

\title{Tidal wave in $^{102}$Pd: \\Rotating condensate of up to seven d-bosons}

\author{ S. FRAUENDORF$^*$, M. A. CAPRIO, and J. SUN }

\address{Department of Physics, University Notre Dame, IN  46557, USA\\
$^*$E-mail: sfrauend@nd.edu}

\begin{abstract}
The yrast states of even even vibrational and transitional nuclei are interpreted as a rotating condensate
 of interacting d-bosons and the corresponding semi-classical tidal wave concept. A simple experimental manifestation
  of the  anharmonicity  caused by the boson interaction is found. The interpretation is substantiated by
  calculations based on the Collective Model and the Cranking Model. 
  
\end{abstract}


\bodymatter
\vspace*{1cm}
\section{Were to find the states with highest number of phonons?} 
\begin{figure}[t]
\begin{minipage}[t]{0.495\linewidth}
\centering
\psfig{file=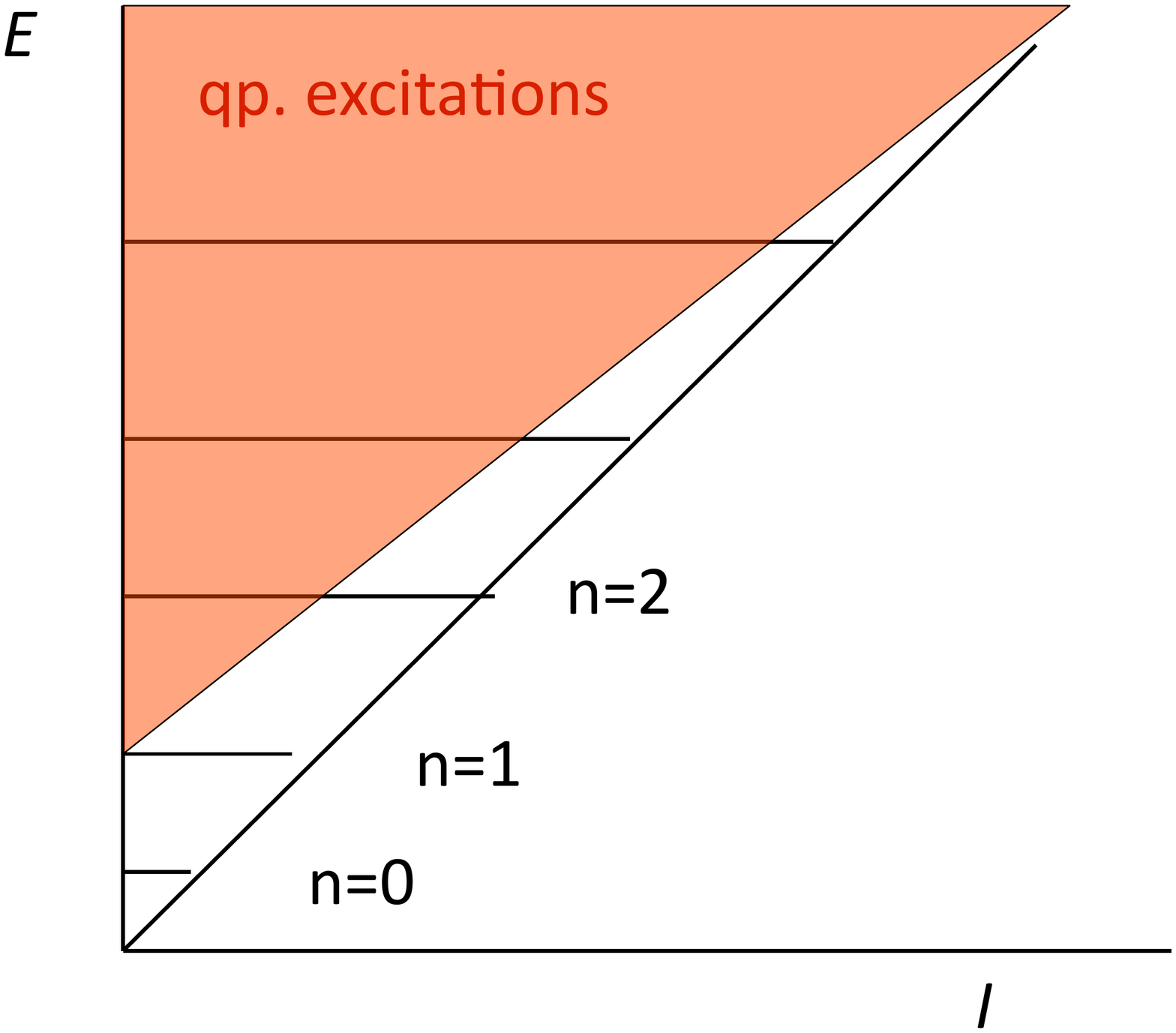,width=\linewidth}
\caption{Schematic excitation spectrum of a vibrational nucleus displayed 
for different angular momenta $I$. The states of minimal $E$ for given $I$ 
are indicated as the yrast line to the right. }
\label{f:vib}       
\end{minipage}
\hfill
\begin{minipage}{0.495\linewidth}
\centering
\vspace*{-1cm}
\psfig{file=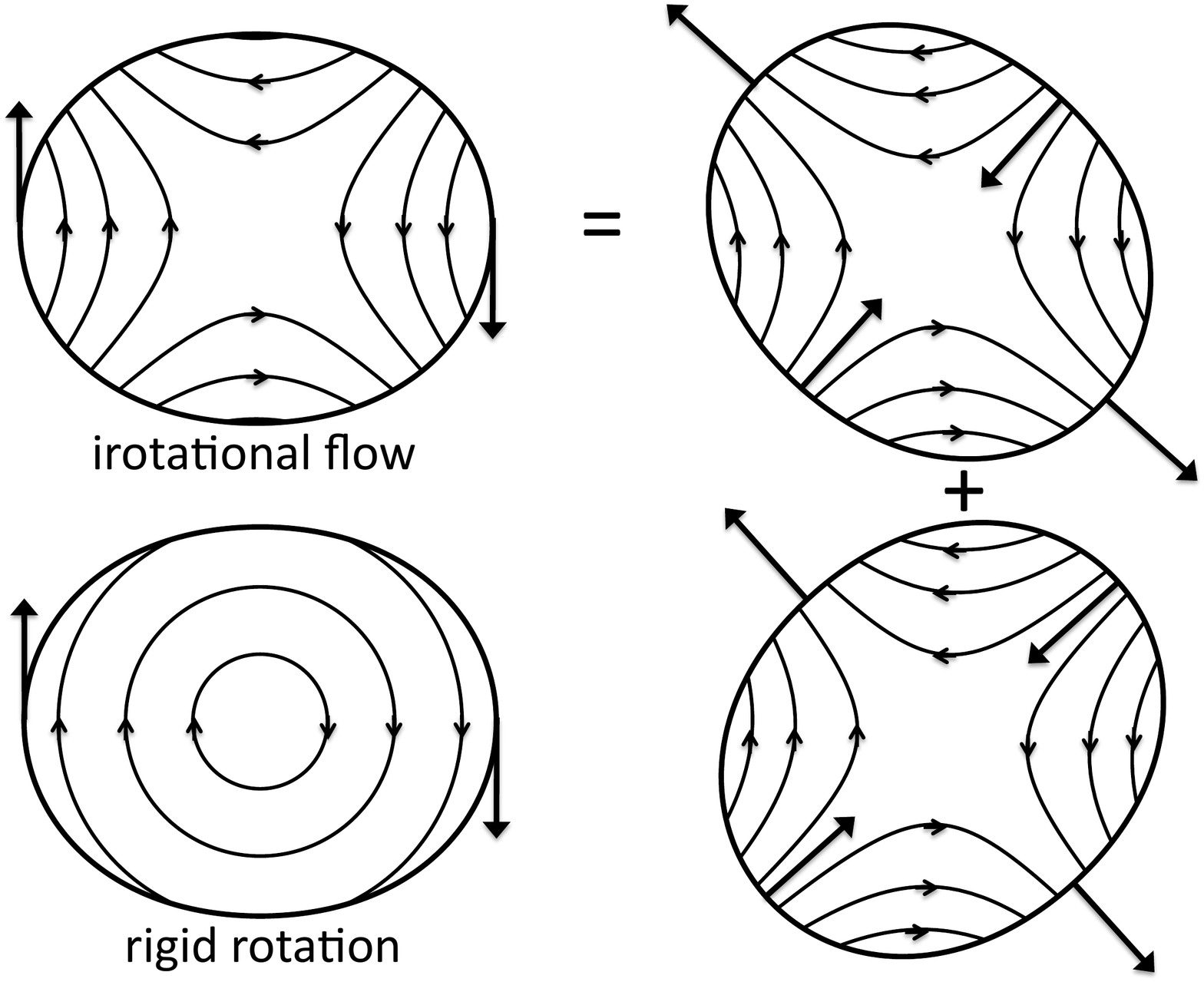,width=\linewidth}
\caption{Flow pattern of a wave running  about the surface of a drop of ideal
liquid (upper left), standing waves  (pulsating vibrations) of such a droplet (right),  
and a rigidly rotating body (lower left). The running (tidal) wave can be thought as a superposition
of two standing waves pulsating with a phase shift of $\pi/2$. From Ref. \cite{Frau10}~. }
\label{f:flow} 
\end{minipage}
\end{figure}

The collective quadrupole excitations of nuclei are classified as "rotational" and "vibrational" with a rigid rotor
and a harmonic vibrator being the limiting ideal cases. Rotational bands that extend over ten and more states
are ubiquitous.  Vibrational excitations of spherical nuclei are much less distinct. While the two-phonon triplet is often
observed, identification of all members of the three-phonon multiplet is already problematic. 
 Fig. \ref{f:vib} schematically shows the location of  
 the vibrational states  and of the quasiparticle excitations for a spherical nucleus. With increasing phonon number $n$, the collective states are embedded into a progressively dense background of quasiparticle excitations. The coupling  to the quasiparticle background 
fragments the collective states, which cease to exist as individual quantum states. The density of quasiparticle excitations 
is lowest near the yrast line, which is the sequence of states with minimal energy for a given angular momentum $I$. 
With increasing phonon number $n$,
the yrast members of the vibrational multiplets keep their identity as collective quantum states longest. 
Moreover, they couple to high-j two-quasiparticle excitations, which have a simple structure.  Hence, the 
vibrational states with highest
phonon number are expected at the yrast line. In this talk, we discuss $^{102}_{46}$Pd$_{56}$. 
The recent lifetime measurements by A. D. Ayangeakaa and U. Garg in collaboration with the ANL group \cite{lifetimes},  
have provided clear evidence for the identification of the seven-boson yrast state. 

\section{The tidal wave concept}
If quadrupole vibrations are  are contrasted with rotation, one usually refers to the pulsating standing wave shown on
the right hand side of Fig. ~\ref{f:flow}. This mode represents the zero-angular momentum members of the vibrational
 multiplets, which are most rapidly drowned in the sea of quasiparticle excitations. The vibrational
 yrast states correspond to a  wave that travels over the nuclear surface.  
 The surface  rotates with the constant angular velocity $\omega$ as in the case of the rotation of a rigid body.
 However,  the flow pattern is irrotational. As characteristic  for a surface wave,  liquid moves from the wave front under
 the crest to the back side. The name "tidal wave" has been suggested for the yrast mode \cite{Frau10} because 
 of its similarity with tidal waves on the ocean. 
 %There is a difference. Water is viscous, and the tidal waves have the 
 %frequency of the driving gravitational forces of the moon and the sun, as for any driven oscillator. In the case
 %of an undamped oscillator (ideal liquid, nucleus) the wave runs with the natural frequency. 
  The energy and 
 the angular momentum increase with the amplitude of the wave, whereas the frequency stays constant.
 In the case of rigid rotation the energy and the angular momentum increase with the angular frequency while
 the shape remains unchanged.   
 
\begin{figure}[t]
\begin{minipage}[t]{0.495\linewidth}
\begin{center}
\vspace*{-2.2cm}
\psfig{file=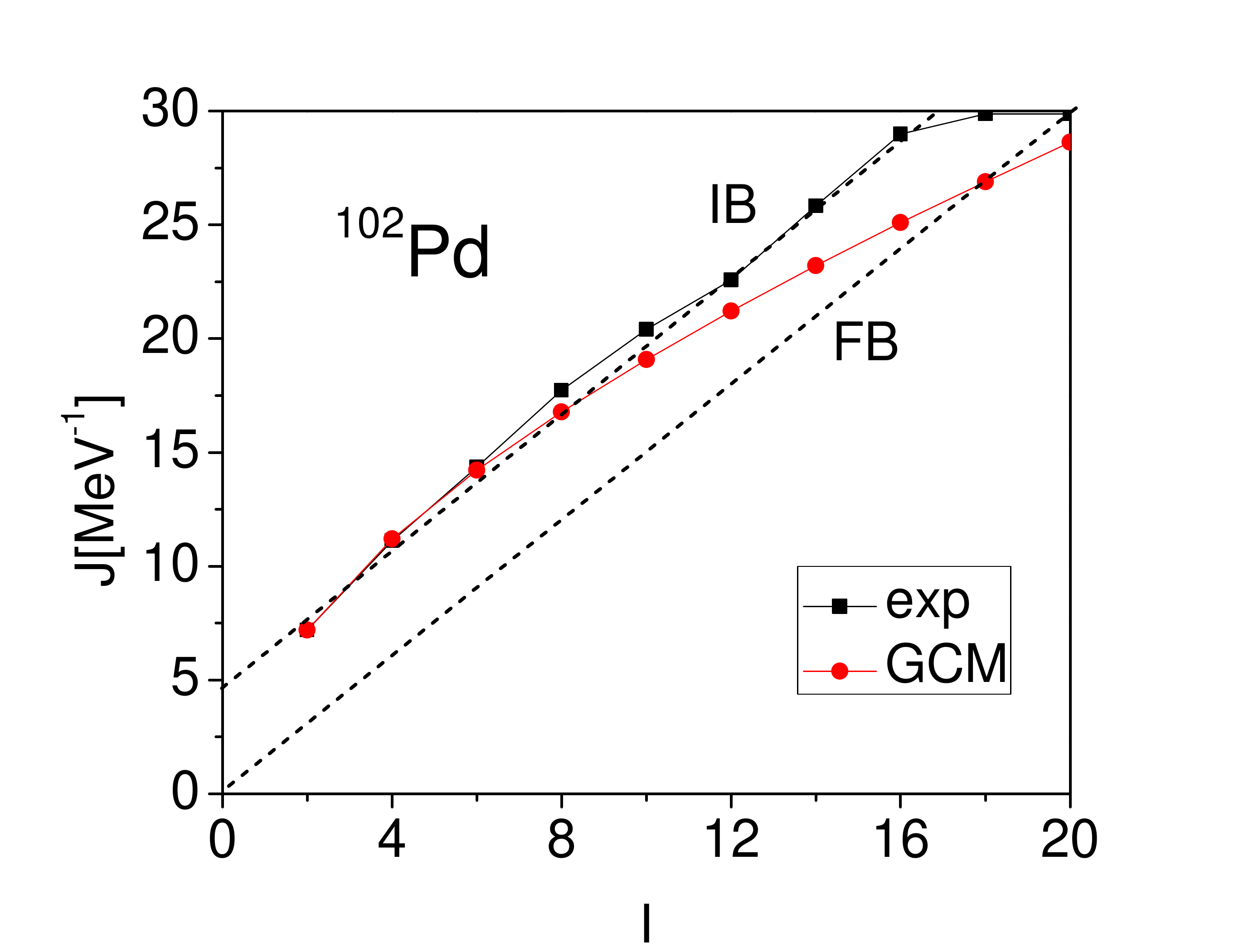,width=\linewidth}
\end{center}
\end{minipage}
\hfill
\begin{minipage}{0.495\linewidth}
\begin{center}
\psfig{file=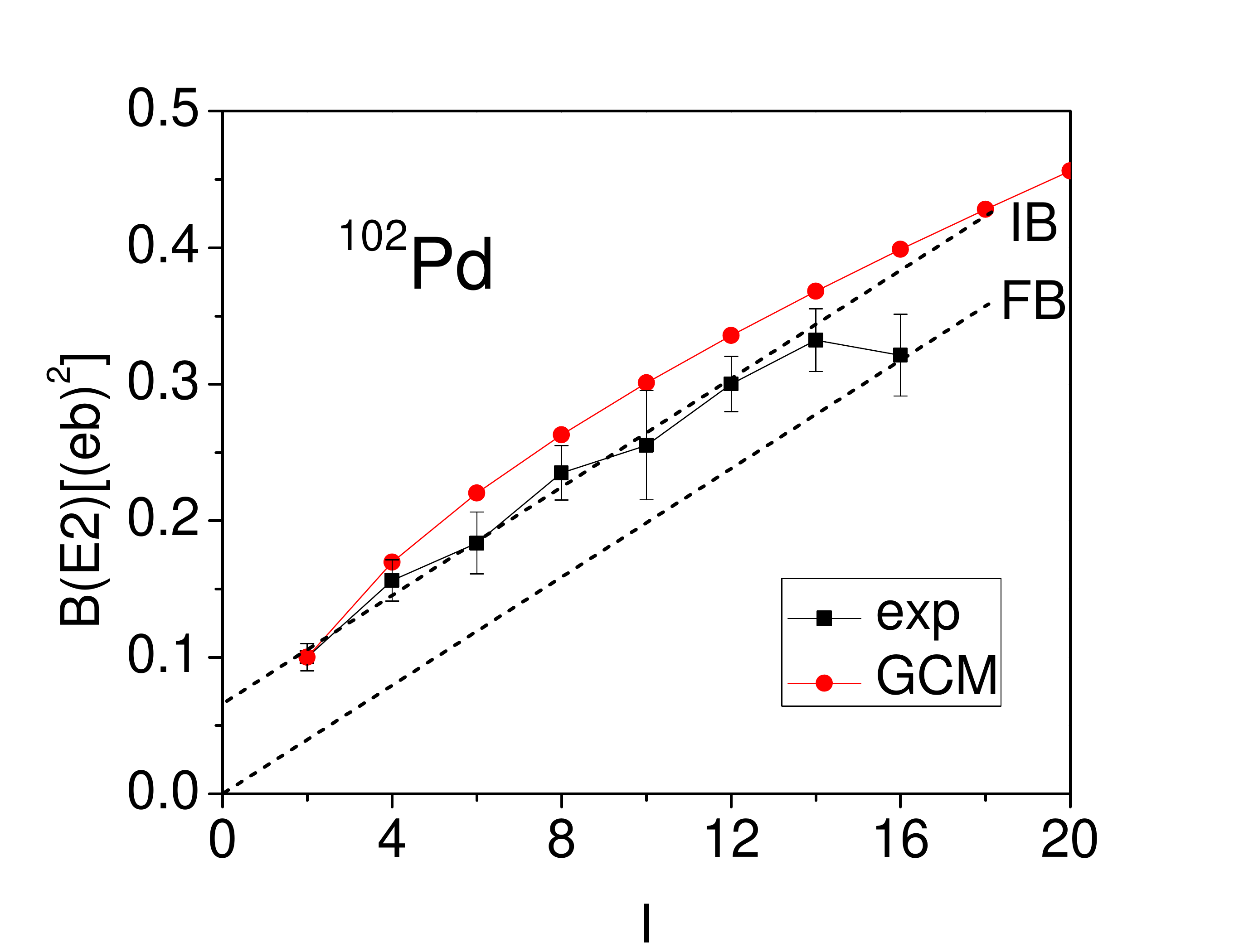,width=\linewidth}
\end{center}
\end{minipage}
\caption{The moment of inertia $J$ (left) and the  $B(E2,I\rightarrow I-2)$ transition probabilities (right)
of the yrast states of $^{102}$Pd. The dashed line FB (free bosons)
shows the limit of harmonic bosons. The dashed line IB (interacting bosons) illustrates the near linear
trend of the interacting bosons.   GCM shows a calculation with the model parameters fitted to the low-lying spectrum. }
\label{f:JIBE2lim}
\end{figure}

The yrast line of vibrational nuclei consists of a sequence of stacked d-bosons, which align their angular momenta.
For sufficient large boson number this  can be considered as a rotating condensate. In the ideal case of $n$ non-interacting bosons one
has
\begin{eqnarray}
E=\Omega\left(n+\frac{5}{2}\right),~~~I=2n,\\
B(E2,I\rightarrow I-2)=\eta^2\Delta \beta^2n =\frac{1}{2}\eta^2\Delta \beta^2I,~~~\eta=\frac{3ZR^2}{4\pi},
\end{eqnarray} 
where $\Omega$ is the frequency of the d-bosons, 
$\beta$ the Bohr deformation parameter of the Generalized Collective Model (GCM) \cite{GCM}, and $\Delta \beta^2$ its
zero point amplitude. For characterizing the energies of the tidal wave it is useful to introduce the angular 
frequency  and the moment of inertia according to the classical relations  $\omega=dE/dI$ and $J=I/\omega$,
respectively, 
\begin{equation}
\omega=\frac{\Delta E}{\Delta I}=\frac{1}{2}\left(E(I)-E(I-2)\right),~~~J=\frac{I}{\omega}.
\end{equation}
In the case of free bosons, $\omega=\Omega/2$ is constant. The moment of inertia  $J$ and the $B(E2)$ are proportional
 to $I$, i. e. their ratio  is $I$-independent.  This limit is indicated by the line FB (free bosons) in Fig. \ref{f:JIBE2lim}. The combination
 $<\beta^2>=\Delta \beta^2n$ has the meaning of an average deformation. Therefore the angular momentum increases, because the moment of inertia increases $\propto <\beta^2>$, while the angular frequency $\omega$ is constant. In the case
 of a rigid rotor, $J$  and  $<\beta^2>$ are constant and $\omega\propto I$. Real nuclei, are between the two limits. It seems
 appropriate to classify the yrast states as  "rotational" when the increase of $I$ is mainly caused by an increase of $\omega$
 (picket fence $E_\gamma$) and "vibrational" when it is mainly caused by an increase of $J$ (roughly constant $E_\gamma$).
 The latter case is the inharmonic tidal wave mode or the rotating condensate of interacting d-bosons.
\section{Phenomenology of the anharmonicity}
Fig. \ref{f:JIBE2lim} demonstrates that the yrast line of $^{102}$Pd classifies as an anharmonic tidal wave. 
The moment of inertia is a nearly linear function of $I$ indicated by the line IB (interacting bosons).
 It deviates from the harmonic limit (FB) by the small offset at $I=0$, which is a measure of the anharmonicity. The $B(E2)$ values behave in the same way, such that
the ratio $B(E2)/J$ is constant within the experimental uncertainties. The GCM  suggests that
both $BE(2)$ and $J$ are $\propto \beta^2$, which implies that their ratio is $I$-independent \cite{GCM}. 
 The available data in vibrational and transitional nuclei in the mass 100 region 
confirm  this observation in a systematic way. Thus, one can easily predict the E2-lifetimes of the yrast levels 
with $I>2$ by the relation 
\begin{equation}
B(E2,I^+\rightarrow(I-2)^+)=B(E2,2^+\rightarrow 0^+)I\frac{ E_\gamma(2^+)}{E_\gamma(I^+)}.
\end{equation}
Employing Grodzin's rule for the 2$^+$ states \cite{raman}
\begin{equation}
\frac{B(E2,2^+\rightarrow 0^+)[(eb)^2]}{E_\gamma[keV]}=2.6(eZ)^2A^{-2/3}
\end{equation}
one can calculate the $B(E2)$ of the yrast states from their transition energies.

The experimental $BE2,I\rightarrow I-2)$ and $J(I)$ values indicate
 that the ground state has some 
deformation $\beta^2(I=0)$ ($>\Delta \beta^2$ )  which increases 
linearly with $I$.
In classical terms, the tidal wave starts with a small deformation and increases its amplitude along the 
yrast line.   In quantum language, the condensate of interacting bosons rotates
like a  condensate of free aligned d-bosons to which a small fraction of non-aligned d-bosons is added,
which generate the $I$-independent part of  $\beta^2(I)$.  The stacking of d-bosons is seen up to $n=7$,
i.e. $I=14$ in the left panel of Fig. \ref{f:JIBE2lim}. The $B(E2,16_1^+\rightarrow 14^+_1)$ belongs to the transition
that connects  the 16$_1^+$ two-quasiparticle state with the 14$^+_1$ seven-boson state. For $I\ge 16$ the
s-configuration composed of two aligned h$_{11/2}$ quasineutrons takes over the yrast line. 
 For $I\ge 16$, the moment of inertia $J(I)$ in the left panel is calculated from the yrare states $I^+_2$.  The 
16$_2^+$ state is most likely the eight-boson state.  The saturation of the curve for $I\ge 18$
indicates that the yrare states  also become two-quasiparticle excitations.  

\section{Description by means of collective models}

We investigated the anharmonicity of the tidal wave  by means of the GCM version of Ref. \cite{caprio}, 
which involves, in addition to the quadratic, a cubic and a quartic potential. That is, the  interaction between the bosons
is a combination of third and fourth order terms. Ref. \cite{caprio} adjusted the potential parameters  to the energy ratios
 between the lowest $0^+_2$, 2$^+_1$, 2$^+_2$, 3$^+_1$, and 4$^+_1$ collective quadrupole excitations. 
 A good description of the of the 
 relative energies and $B(E2)$ ratios was obtained. Fig. \ref{f:JIBE2lim} shows how the same GCM calculation describes
  the tidal wave, where the energies and the $B(E2)$ are scaled to  the experimental values for the 2$^+_1$ state.  
 The GCM reasonably well reproduces the slope and offset of the experimental curves. The ratio 
 $B(E2,I\rightarrow I-2)/J(I)$ is nearly $I$ independent. However, the calculation deviates from the 
 linear behavior of the experimental curves. This may indicate that the quartic form of the Bohr Hamiltonian is not quite
 appropriate.     
 
 We also attempted describing the tidal wave by means of the Interacting Boson Model (IBM 1) using  parameters determined
  in the same way as for the GCM. IBM 1 fails to describe the multi-phonon yrast states.
 The function $J(I)$ is found to be decreasing, and the $B(E2)$ values decrease for $I>6$, which is
 in striking contrast to experiment.  The reason for the discrepancy is the cut off in boson number.
 According to IBM 1 counting, the maximal boson number is 5 for $Z=46$ and $N=56$, 
 whereas, as discussed above, the experiment indicates the presence of the  seven- and, tentatively,
 eight-boson states.

\section{Microscopic mean field description }
\begin{figure}[t]
\begin{minipage}[t]{0.495\linewidth}
\begin{center}
\vspace*{-2.2cm}
\psfig{file=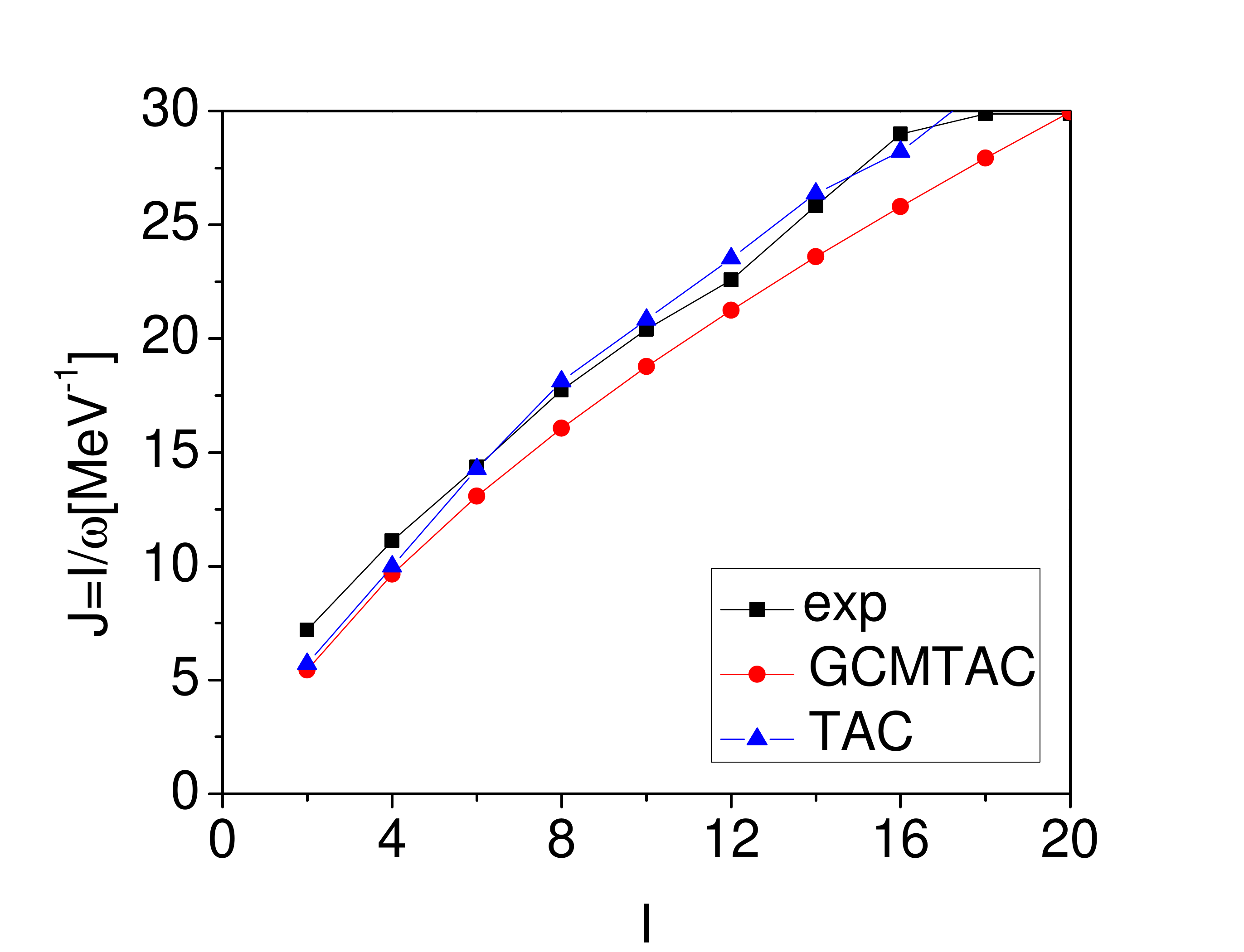,width=\linewidth}
\end{center}
\end{minipage}
\hfill
\begin{minipage}{0.495\linewidth}
\begin{center}
\psfig{file=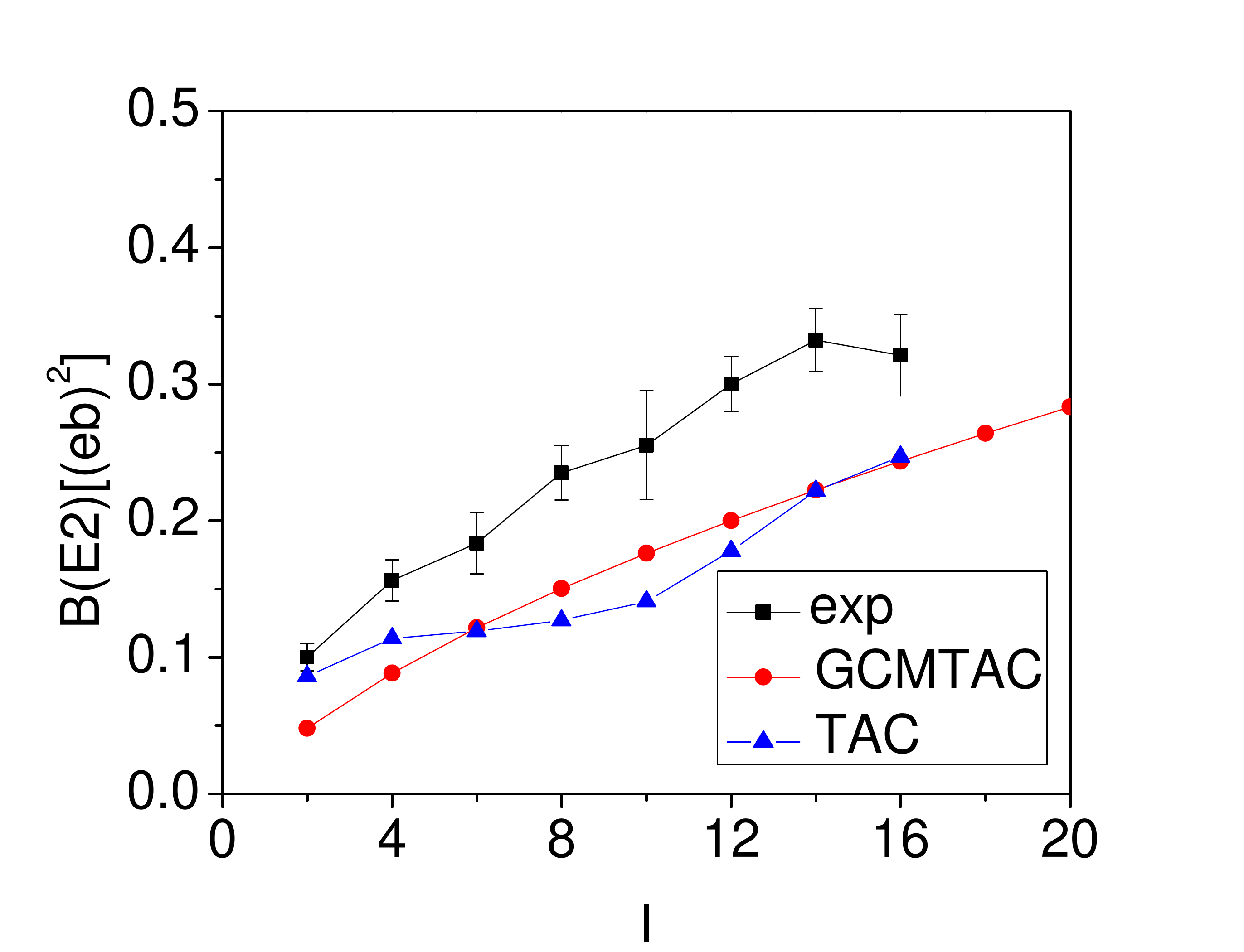,width=\linewidth}
\end{center}
\end{minipage}
\caption{The moment of inertia $J$ (left) and the  $B(E2,I\rightarrow I-2)$ transition probabilities (right)
of the yrast states of $^{102}$Pd. GCMTAC  shows the GCM calculation with the quartic potential fitted
to the microscopic TAC PES and the mass parameter adjusted to experiment. TAC is obtained by minimizing
the microscopic energy with respect to the deformation parameters $\beta$ and $\gamma$.    }
\label{f:JIBE2TAC}
\end{figure}

The tidal wave has a static deformed shape in the co-rotating frame of reference. This leads to the microscopic
description suggested in Ref. \cite{Frau10}, which is based on the rotating mean field.
 Ref. \cite{Frau10} applied the SCTAC (shell correction tilted axis cranking) version \cite{qptac} of 
the Cranking Model to even-even nuclides with $44\le Z\le 48$ and $56\le N\le 66$. SCTAC calculates the energy
for given expectation value of the angular momentum operator  equal to $I$  by means of the micro-macro method using a deformed Woods-Saxon potential. The energy is minimized with respect to the deformation 
parameters $\beta$ and $\gamma$. Deformed solutions were found for $I\ge 2$ even when 
the solution was spherical for $I=0$.
These calculations describe the collective yrast states rather well. They also 
describe the intrusion of the aligned h$_{11/2}$ two quasi neutron states into the yrast line, which 
causes the back bending phenomenon seen in most of the studied nuclei.  Figs. \ref{f:JIBE2TAC}
and \ref{f:110cdBE2TAC} shows two examples,  which we refer to as TAC. In the case of $^{102}$Pd, 
the function $J(I)$ is very well reproduced.
The $B(E2)$ values show the characteristic increase with  $I$. However, their absolute values is too small.
Note, there are no adjustable parameters in the calculation. Of course, one could achieve good agreement with 
experiment by introducing an effective charge.
In the case of $^{110}$Cd, the TAC function $J(I)$  (not shown) agrees very well with experiment.
At $I=12$, the aligned h$_{11/2}$ two-quasi neutron configuration
becomes yrast, which is very accurately reproduced by the TAC calculation. 
As seen in Fig. \ref{f:110cdBE2TAC}, the TAC calculation underestimates the $(BE2)$ values of the tidal wave branch
$I\le 10$.  The drop at $I=12$ indicates the change to the   h$_{11/2}$ two-quasineutron configuration, the
$B(E2)$ values of which are well reproduced.   The TAC calculations of Ref. \cite{Frau10} 
give also too small $B(E2)$ values in the tidal branches of $^{104}$Ru and $^{106-110}$Pd, for which lifetimes
have been measured up to $I=8$.

\section{Mixed description}
\begin{figure}[t]
\begin{minipage}[t]{0.49\linewidth}
\begin{center}
\vspace*{-3.5cm}
\psfig{file=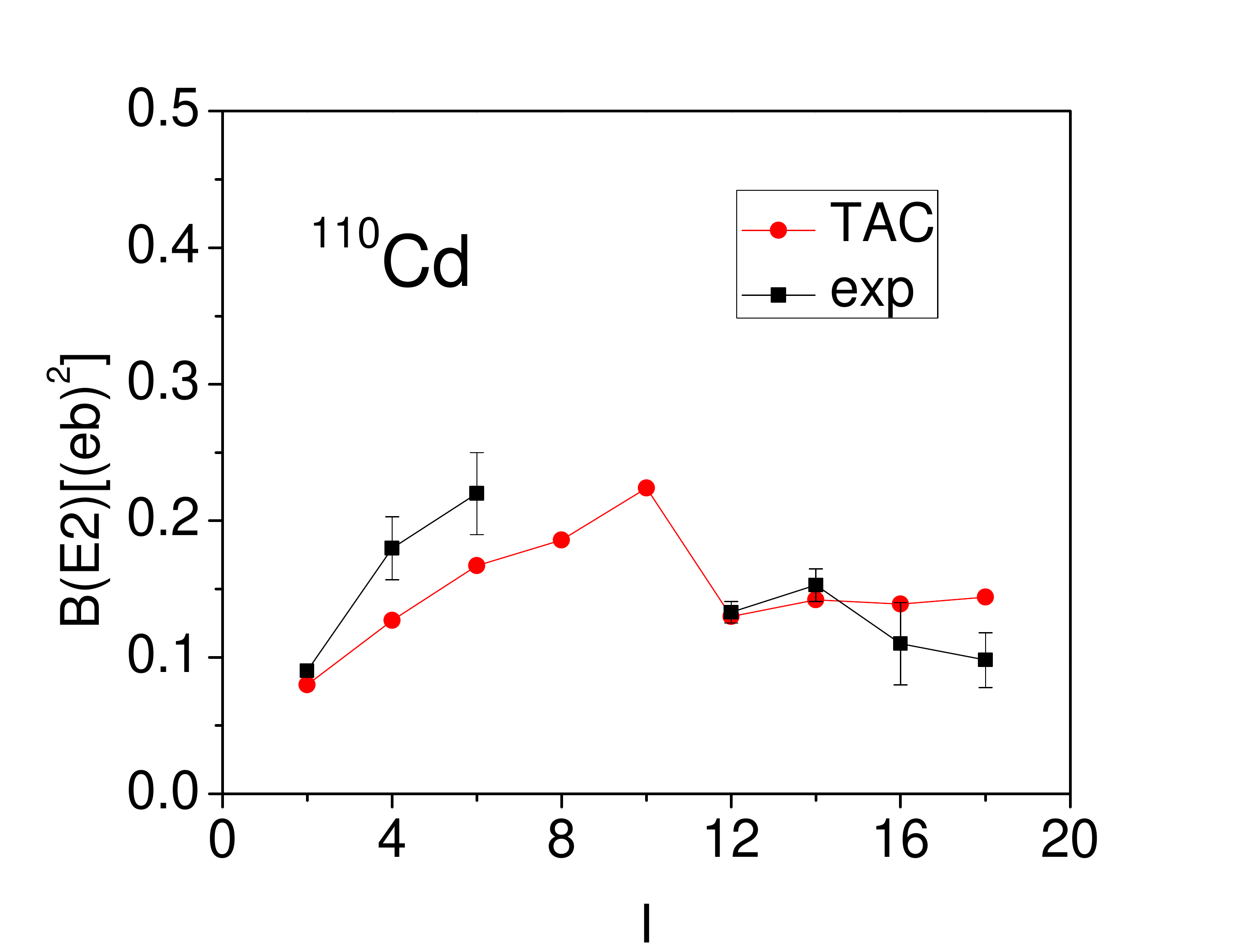,width=\linewidth}
\end{center}
\vspace*{-0.07cm}
\caption{ The $B(E2,I\rightarrow I-2)$ transition probabilities 
of the yrast states of $^{110}$Cd.  TAC is obtained by minimizing
the microscopic energy with respect to the deformation parameters $\beta$ and $\gamma$.  }
\label{f:110cdBE2TAC}
\end{minipage}
\hfill
\begin{minipage}{0.49\linewidth}
\begin{center}
\psfig{file=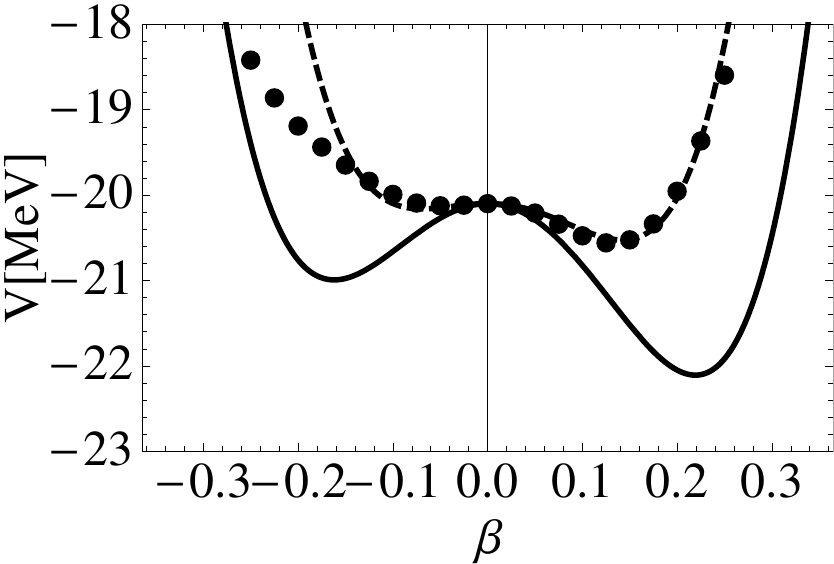,width=\linewidth}
\end{center}
\caption{Ground state potentials of $^{102}$Pd. Dots: microscopic TAC potential,  dashed curve: quartic GCM potential fitted 
to the dots, full curve: phenomenological GCM potential.    }
\label{f:pots}
\end{minipage}
\end{figure}

The TAC calculations treat the quadrupole degree of freedom in a classical way, which disregards
zero point fluctuations. In order to take them into account, we fitted the quartic potential of GCM Hamiltonian to 
the $I=0$ potential energy surface (PES) calculated by means of TAC. Fig. \ref{f:pots} shows the fit for axial shapes.   
The mass parameter $B=90\hbar^2$MeV$^{-1}$ of the GCM was adjusted to the experimental energy scale. 
The  factor $\eta$ in Eq. (2) was used to connect the deformation parameter $\beta$
with the  $B(E2)$. This relation  is very close to the microscopic relation
in the TAC calculations.  
The results obtained with this GCM Hamiltonian are denoted by GCMTAC in Fig. \ref{f:JIBE2TAC}. The 
function $J(I)$ is well reproduced. The $B(E2)$ values are too small. It is surprising that results of the  TAC and
the GCMTAC calculations are very similar, that is, that the inclusion of the zero point fluctuations seems not to 
lead to dramatic changes. 

The small $B(E2)$ values of the TAC and GCMTAC calculations can be traced to the micro-macro TAC 
PES. As seen in Fig. \ref{f:pots}, the potential of the phenomenological GCM Hamiltonian, 
which reproduces the $B(E2)$ well,  is much wider.  
We investigated a number of parameter sets from the literature for the Woods-Saxon potential used 
 in the SCTAC calculations. The resulting PES were similar to the one in Fig. \ref{f:pots} and the 
 $B(E2)$ values too small.
 Adopting the Nilsson potential did not resolve the problem.    On the other hand, the agreement 
 of the calculated $B(E2)$ values with experiment in $^{110}$Cd for $I\ge 12$ gives credibility to the 
 TAC PES. The CNS model \cite{CNS} and the TRS calculations  \cite{TRS} 
   reproduce  energies and $B(E2)$ values of high spin states of many 
 nuclei, including the discussed ones. CNS and TRS apply the micro-macro method 
 based on the Nilsson and Woods-Saxon potentials, which gives additional strong credibility to the TAC PES. 
 At this point, it remains unclear why the microscopic approach underestimates the $B(E2)$ values of the tidal wave.         

\section{Summary}
The states of highest d-boson number will be found along the yrast line. The seven-phonon yrast state has been
 identified in $^{102}$Pd.  The semi-classical concept of a tidal wave running over the nuclear surface
characterizes the nature of the d-bosonic yrast states. The boson interaction generates anharmonicity, which 
shows up as a constant shift of the $B(E2,I\rightarrow I-2)$ value and the moment of inertia $J(I)$ as functions of the spin.  
The ratio  $B(E2,I\rightarrow I-2)/J(I)$ is found to be approximately $I$-independent, which allows one to predict the
$B(E2)$ values from the yrast energies. A collective
 Hamiltonian with a quartic potential, the parameters of which were adjusted to
low-lying collective states, accounts well for the experimental energies and $B(E2)$ values in $^{102}$Pd.
Microscopic calculations without adjustable parameters, based on the Cranking Model, reproduce the energies, but
underestimate the $B(E2)$ values by a factor of about 2/3.  

\section*{Acknowledgments}
Supported by the DoE Grant DE-FG02-95ER4093. We thank A. D. Ayangeakaa and U. Garg for making the lifetime data available to us.

\bibliographystyle{ws-procs9x6}
\bibliography{ws-pro-sample}

\end{document}